\documentclass[12pt]{iopart}

\usepackage{bm}
\usepackage{graphicx}
\usepackage{subfigure}
\usepackage{color}
\begin{document}

\title[DMRG-ENPT2]{Multi-Reference Epstein-Nesbet Perturbation Theory with Density Matrix Renormalization
Group Reference Wavefunction}

\author{Yinxuan Song$^1$, Yifan Cheng$^1$, Yingjin Ma$^{2,3}$, Haibo Ma$^1$}

\address{$^1$ School of Chemistry and Chemical Engineering, Nanjing University, Nanjing 210023, China}
\address{$^2$ Department of High Performance Computing Technology and Application Development, Computer Network Information Center, Chinese Academy of Sciences, Beijing 100190, China
}
\address{$^3$ Center of Scientific Computing Applications \& Research, Chinese Academy of Sciences, Beijing 100190, China
}
\ead{yingjin.ma@sccas.cn (Y.M.) or haibo@nju.edu.cn (H.M.)}
\vspace{10pt}
\begin{indented}
\item[]December 2019
\end{indented}

\begin{abstract}
The accurate electronic structure calculation for strongly correlated chemical systems requires an adequate description for both static and dynamic electron correlation, and is a persistent challenge for quantum chemistry. In order to account for static and dynamic electron correlations accurately and efficiently, in this work we propose a new method by integrating the density matrix renormalization group (DMRG) method and multi-reference second-order Epstein-Nesbet perturbation theory (ENPT2) with a selected configuration interaction (SCI) approximation. Compared with previous DMRG-based dynamic correlation methods, the DMRG-ENPT2 method extends the range of applicability, allowing us to efficiently calculate systems with very large active space beyond 30 orbitals. We demonstrate this by performing calculations on H$_2$S with an active space of (16e, 15o), hexacene with an active space of (26e, 26o) and 2D H$_{64}$ square lattice with an active space of (42e, 42o).
\end{abstract}

%\maketitle
\section{Introduction}
In recent years, the density matrix renormalization group (DMRG) method \cite{white1992density, white1992real, mitrushenkov2001quantum, chan2002highly, legeza2003controlling, legeza2003qc, legeza2003optimizing, legeza2004quantum, chan2004algorithm, moritz2005convergence, moritz2005relativistic, rissler2006measuring, legeza2008applications, chan2009density, marti2010density, chan2011density, ma2013assessment, legeza2015advanced, chan2016matrix,shuai1997hubbardpeierls, ren2016ispt} has been shown to be a prominent new quantum chemical approach to approximate the full configuration interaction (FCI) solution within a large active space with only polynomial computational costs.\cite{chan2002highly} However, handling electron correlation in realistic chemical systems is far more complicated than that in restricted active space calculation, and dynamic correlation outside the active space needs to be considered in order to get quantitative results. Over the last few years, a number of methods have been proposed to achieve this goal, including DMRG-canonical transformation (CT) \cite{yanai2010multireference}, DMRG-complete active space second-order perturbation
theory (CASPT2) \cite{kurashige2011second, kurashige2014caspt2}, DMRG-\emph{N}-electron valence perturbation
theory (NEVPT2) \cite{guo2016n, roemelt2016projected, freitag2017multireference}, DMRG-multi-reference configuration interaction (MRCI) \cite{saitow2013multireference, saitow2015fully, luo2018externally}, and DMRG-tailored coupled cluster (TCC) \cite{veis2016coupled, faulstich2019numerical}, matrix product state perturbation theory (MPSPT) \cite{sharma2014mpspt,sharma2017icmpspt}. Because of the too huge number of the reference configurations within a very large active space in DMRG calculation, usually internally contraction (ic) \cite{werner1982self, werner1988efficient} or external contraction (ec) \cite{siegbahn1980direct} approximations and/or a truncation for reference configurations have to be adopted in these post-DMRG dynamic correlation calculations. {Alternatively, time-dependent formulation of multi-reference perturbation theory was proposed to incorporate dynamic correlation for reference wavefunctions with up to 24 active electrons and orbitals without computation of the four-particle reduced density matrix \cite{sokolov2016tdMRPT,sokolov2017tdNEVPT}.} Hybridizations with density functional theory (DFT) \cite{hedegard2015dmrgRangeSeparation} or pair density functional theory (PDFT) \cite{sharma2019pdft} were also implemented as post-DMRG treatments. Recently, we have shown that DMRG-ec-MRCI is capable of adequately describing the static and dynamic electron correlation in systems with large active spaces beyond 30 orbitals, e.g. chromium dimer (Cr$_2$) with an active space of (12e, 42o), oligocenes with active spaces up to (38e, 38o) and Eu-BTBP(NO$_3$)$_3$ complex with an active space of (38e, 36o). \cite{luo2018externally} However, for the purpose of dealing with even larger active spaces, the development of economic treatments for dynamic electron correlations on top of DMRG calculations is still highly necessary.

Perturbation theory (PT) offers a simple and efficient way among various dynamic correlation treatments. This type of approximation goes as a power series of the perturbation parameter $\epsilon$ defined by $\hat{H}=\hat{H}_0+\epsilon\hat{V}$, where $\hat{H}_0$ is the zeroth-order Hamiltonian operator while $\hat{V}$ is the perturbation.
In the domain of multi-reference PT (MRPT), CASPT2 \cite{andersson1990second, andersson1992second} and NEVPT2 \cite{angeli2001introduction} are widely used, with a difference in defining $\hat{H}_0$. In CASPT2 \cite{andersson1990second, andersson1992second}, the zeroth-order Hamiltonian is defined in terms of a Fock-type one-electron operator (the generalized Fock operator), however it is well-known that CASPT2 suffers from a few defects: the energy of systems with open shells will be too low, and it faces intruder states due to too small denominators. \cite{ghigo2004modified}
In order to address the two-electron interaction effect in the zeroth-order wavefunction, Dyall \cite{dyall1995choice} suggested an auxiliary two-electron zeroth-order Hamiltonian, which is equivalent to the full Hamiltonian within the complete active space (CAS). Based on Dyall's Hamiltonian, Malrieu \textit{et al.} \cite{angeli2001introduction} proposed NEVPT2 as an alternative MRPT method, which is strictly additive and free of intruder states. Besides CASPT2 and NEVPT2, Epstein-Nesbet PT (ENPT) \cite{epstein1926stark, nesbet1955configuration} can provide a computationally simpler MRPT solution by partitioning the the full configurational space into a variational space and an outer space. The simplicity of only diagonal elements in the outer space for the zeroth-order Hamiltonian in ENPT makes it an attractive alternative to other MRPT methods, since no diagonalization or solving linear equations is required.\cite{murphy1992enpt} Such kind of ENPT2 treatment has been successfully adopted for a quick estimation of the perturbative energy correction in CI by perturbation with multi-configurational zeroth-order wavefunction selected by iterative process (CIPSI) \cite{huron1973iterative}, heat-bath CI (SHCI) \cite{sharma2017semistochastic} and quantum Monte Carlo (QMC) \cite{blunt2018communication, dash2018perturbatively} etc.

Sharma \cite{sharma2018stochasticPT} and also Chan and co-workers \cite{guo2018perturbative, guo2018communication} recently applied ENPT2 corrections for DMRG calculations with a small bond dimension $M$ in the context of matrix-product states (MPS) to approach the DMRG calculations with a large bond dimension $M$ within a given active space. Their works showed that the selected CI-then-perturbation strategy has the capacity to describe the transition metal complex up to tens of active electrons/orbitals. When combining with a good choice of zeroth-order Hamiltonian, the selected CI-then-perturbation strategy can provide highly accurate total energies for challenging systems with significantly reduced computational resources when comparing with the deterministic Epstein-Nesbet perturbed DMRG, and both of stochastic and deterministic algorithms are much cheaper than the original variational DMRG, in large orbital spaces with a mix of correlation strengths.
%and the perturbative DMRG can be used as a cheaper alternative to variational DMRG to estimate an exact ground state energy when nearly hundreds orbitals are employed.

In this work, we integrate DMRG and ENPT2 based on our entanglement-driving genetic algorithm (EDGA) scheme, to describe the static and dynamic electron correlation within and outside the given active space adequately. It demonstrates that the EDGA-based DMRG-ENPT2 approach provides an efficient tool for describing the complex electronic structure of strongly correlated chemical molecules with very large active spaces beyond 30-40 orbitals.

\section{Methodology}

As details of quantum-chemical DMRG have been discussed elsewhere \cite{chan2002highly, legeza2003controlling, chan2009density, marti2010density, chan2011density, ghosh2008orbital, zgid2008density, ma2017scf, mcculloch2007density, schollwock2011density, Keller2016Spin}, herein we only
briefly introduce ENPT2 and selected CI (SCI) as well as how they are employed in the context of DMRG-ENPT2.

In MR-ENPT, the full configurational space is partitioned into a variational space, $\Pi$, spanned by determinants labeled $|D_i\rangle$ and $|D_j\rangle$, and the rest of the space, spanned by determinants labeled $|D_a\rangle$. The zeroth-order Hamiltonian consists of the full Hamiltonian block within $\Pi$ and only the diagonal elements of $\hat{H}$ outside $\Pi$, by defining
\begin{equation}\label{H0}
\hat{H}_0=\sum_{ij\in\Pi}\langle D_i|\hat{H}|D_j\rangle|D_i\rangle\langle D_j|+\sum_{a\notin\Pi}\langle D_a|\hat{H}|D_a\rangle|D_a\rangle\langle D_a|.
\end{equation}
The zeroth-order wave function $|\Psi_0\rangle=\sum_{i\in\Pi}c_i|D_i\rangle$ and the zeroth-order energy $E_0$ are the eigenvector and eigenvalue of $\hat{H}_0$. By virtue of using standard perturbation theory, the second-order energy correction can be calculated by
\begin{equation}\label{ENPT2}
\Delta E_{\rm ENPT2}=\sum_{a\notin\Pi}\frac{{(\sum_{i\in\Pi}\langle D_a|\hat{H}|D_i\rangle c_i)}^2}{E_0-\langle D_a|\hat{H}|D_a\rangle}.
\end{equation}

Next we show how DMRG provide $|\Psi_0\rangle$ for ENPT2 calculation. DMRG wave function (spanned by $L$ orbitals) is usually represented in MPS formulation by
\begin{equation}\label{MPS}
    |\Psi\rangle= \sum_{\sigma_1,...,\sigma_L}\sum_{M_1,...,M_{L-1}}A_{1,M_1}^{\sigma_1}A_{M_1,M_2}^{\sigma_2}...A_{M_{L-1},1}^{\sigma_L}|\sigma_1...\sigma_L\rangle \\
    = \sum_{\bm{\sigma}}A^{\sigma_1}A^{\sigma_2}...A^{\sigma_L}|\bm{\sigma}\rangle,
\end{equation}
where the basis states $|\sigma_l\rangle$ for the $l$-th orbital has four possible occupation status as $\left|\uparrow\downarrow\right>$, $\left|\uparrow\right>$, $\left|\downarrow\right>$ and $\left|0\right>$, $M_{l-1}\times M_{l}$-dimensional matrices
$A^{\sigma_l} = \{A^{\sigma_l}_{M_{l-1},M_l}\}$ are obtained by successive singular value decomposition (SVD) procedures in DMRG sweeps by ignoring the configurations with very small singular values. Collapsing the summation over the $a_l$\ indices as matrix-matrix multiplications results in the last equality. Notice that $m_{l-1}\times m_{l}$-dimensional matrices with the first matrix is $1\times M_1$-dimensional row vector and the last one is $M_{L-1}\times 1$-dimensional column vector, respectively.

%In 2007, Moritz et al. \cite{moritz2007decomposition} rationalized that
The MPS representation for the wave function of \Eref{MPS} can be equivalent to a FCI or CASCI expansion with a Slater determinant (SD) configurational basis ($|\Psi\rangle=\sum_{\sigma_1,...,\sigma_L}c_{\sigma_1...\sigma_L}|\sigma_1...\sigma_L\rangle$) by calculating the CI coefficient $c_{\sigma_1...\sigma_L}$ by
\begin{equation}
    c_{\sigma_1...\sigma_L} = A^{\sigma_1}A^{\sigma2}... A^{\sigma_L}.
\end{equation}
where $A$ matrices for basis transformations can be obtained and kept in DMRG sweeps. This was first rationalized by Moritz \textit{et al.} and can be used in wave function analysis \cite{moritz2007decomposition}.
However, the FCI expansion for a DMRG wave function in the large active space with more than 20 active orbitals would be prohibitive due to the number of SDs would be easily larger than $10^{10}$. \cite{moritz2007decomposition} Two different schemes for efficiently sampling the most important configurations are recently proposed, Monte Carlo based sampling-reconstructed CAS (SR-CAS) algorithm by Boguslawski \textit{et al.} \cite{boguslawski2011construction} and the EDGA proposed by some of us \cite{luo2017efficient}.
Our recent work \cite{luo2018externally} illustrated that it is possible to use a limited number (e.g. thousands, or tens of thousands) of most important configurations that obtained by EDGA to achieve a wave function completeness of 0.99 for large active spaces. We use the completeness (in the range between 0 and 1) to measure the importance proportion of the sampled leading configurations in a full reference wavefunction. It is calculated by the summation of squared coefficients of all selected configurations. The basic idea of EDGA is to sample the most important CASCI-type configurations in a DMRG/MPS wavefunction via exploring a huge determinant configurational space by using a genetic algorithm, in which the orbital entanglements is used to increase the evolution efficiency. We further perform SCI calculations by constructing the Hamiltonian with these selected small number ($10^2\sim10^5$) of important configurations as the basis.
The solution of a SCI calculation with the sampled important configurations will be then used for the zeroth-order wave function $|\Psi_0\rangle$ and the zeroth-order energy $E_0$ in ENPT2 calculation.

All DMRG-CASCI calculations in this work are implemented using the {\sc QCmaquis} DMRG software package \cite{Knecht2016New, keller2015efficient, Keller2016Spin}. The EDGA, SCI and subsequent ENPT2, ec-MRCI calculations are performed with our in-house code.

\section{Results and discussion}
\subsection{H$_2$S, N$_2$ and benzene dimer}

H$_2$S is a medium sized molecular system that allows us to use DMRG-FCI method as a benchmark reference. First, we adopt DFT to optimize the ground state structure of H$_2$S at the B3LYP/6-31G level by using {\sc GAUSSIAN09} \cite{g09} package. The equilibrium H-S-H angle $\theta=94.06^\circ$ and S-H bond length $r=1.379{\rm \AA}$, respectively, then we set $r=1.7{\rm \AA}$ to get a stretched structure while keeping $\theta$ unchanged. In the following Multi-Configuration (MC)/MR electron correlation calculations, a large atomic natural orbitals basis sets that contracted to quadruple-zeta (ANO-L-VQZP) is used. We froze the 1s atomic orbital of S and defined a (16e, 15o) active space, which contains 2s, 2p, 3s, 3p, 3d  atomic orbitals of S and 1s atomic orbitals of H. Both HF canonical orbitals, localized orbitals (by Pipek-Mezey method \cite{pipek1989fast}) and CASSCF natural orbitals are used for the subsequent MC/MR calculations. The DMRG-FCI (16e, 95o) and DMRG-CASCI/DMRG-CASSCF (16e, 15o) calculations with $M=2000$ are performed. The SCI, ec-MRCISD+Q, and ENPT2 calculations are performed using truncated reference wavefunctions constructed via EDGA framework with CI completeness of 0.999.

\begin{figure}
\centering
\includegraphics[scale=0.6,bb=0 0 800 600]{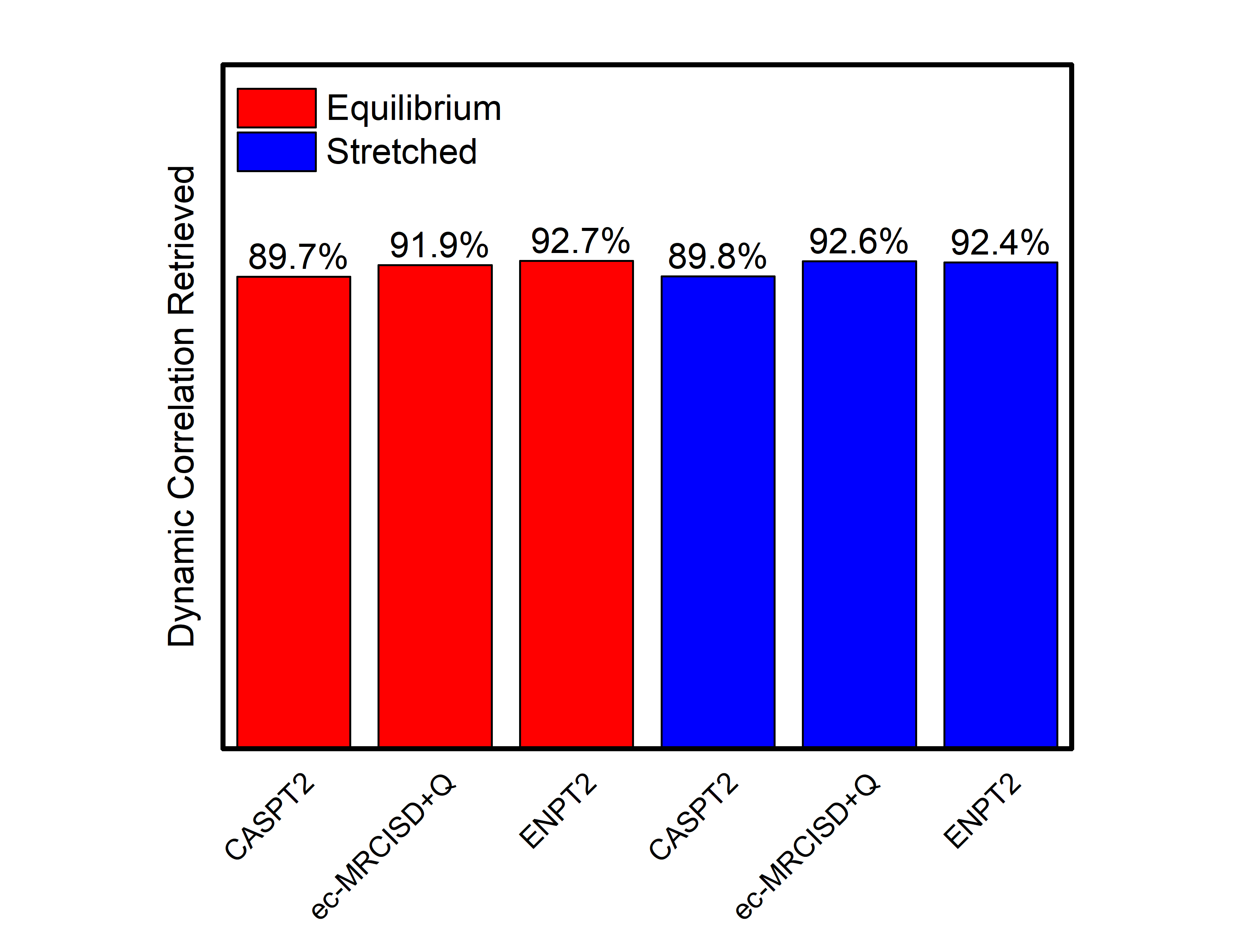}
\caption{Dynamic correlation retrieved by different MR methods using CASSCF orbitals with a (16e, 15o) active space for H$_2$S.}\label{figure1}
\end{figure}

The calculated results are listed in \Tref{table1: h2s}.
It can be found that the results of SCI differs from CASCI around 1$\sim$8 mHartree for different orbitals. However, the number of reference configurations for SCI is about 200 in canonical orbitals and 3000 in CASSCF orbitals, among the full reference configurations of $10^{10}$ for the CASCI/CASSCF wavefunction. It means that we only need a limited number of reference states to get results close to CASCI/CASSCF.
Compared with that of DMRG-FCI in CASSCF orbitals, all of the CASPT2, ec-MRCISD+Q, and ENPT2 can achieve semi-quantitatively satisfactory results and retrieve at least 89.7\%, 91.9\%, and 92.4\% of the total dynamic correlation energies (-146 mHartree for equilibrium H$_2$S and -136 mHartree for the stretched one), respectively, shown in \Fref{figure1}.The results of CASSCF orbitals are closer to the results of DMRG-FCI. This is because CASSCF calculations already account for part of the correlation between the initial active orbitals and the outer ones to some extent, through the MC orbital optimizations. One may also notice that the dynamic correlation contribute to a significant increase of 15 mHartree in canonical orbitals and 10 mHartree in CASSCF orbitals from DMRG-CASCI(16e, 15o) to DMRG-FCI(16e, 95o) for the energy gap between the energies of the equilibrium and stretched structures.
Both DMRG-ec-MRCISD+Q and DMRG-ENPT2 can reproduce this increase caused by dynamic correlation, and their results are close to each other. It demonstrates that, with a much cheaper computational cost than the variational DMRG and DMRG-MRCI, our DMRG-ENPT2 can still efficiently count the dynamic correlation.

\begin{table}
\centering
\caption{Ground state energies of H$_2$S using different MC/MR methods}\label{table1: h2s}
\lineup
\begin{tabular}{@{}c|ccc}
\br
H$_2$S (16e, 15o)&$E_{equilibrium}$/Hartree&$E_{stretched}$/Hartree&$\Delta$/mHartree\\
\mr
DMRG-FCI(16e, 95o)&\-398.99268&\-398.91913&73.55\\
\mr
&\multicolumn{3}{c}{Canonical orbitals}\\
\mr
DMRG-CASCI&\-398.72456&\-398.66592&58.64\\
DMRG-SCI&\-398.72325&\-398.66438&58.87\\
DMRG-ec-MRCISD+Q&\-398.96250&\-398.89187&70.63\\
DMRG-ENPT2&\-398.98366&\-398.90819&75.47\\
CASPT2&\-398.96176&\-398.88891&72.85\\
\mr
&\multicolumn{3}{c}{Localized orbitals}\\
\mr
DMRG-CASCI&\-398.72456&\-398.66592&58.64\\
DMRG-SCI&\-398.71676&\-398.65868&58.08\\
DMRG-ec-MRCISD+Q&\-398.91496&\-398.85508&59.87\\
DMRG-ENPT2&\-398.82482&\-398.77175&53.07\\
CASPT2&\-398.96176&\-398.88891&72.85\\
\mr
&\multicolumn{3}{c}{CASSCF orbitals}\\
\mr
DMRG-CASCI&\-398.84627&\-398.78301&63.27\\
DMRG-SCI&\-398.84124&\-398.77675&64.50\\
DMRG-ec-MRCISD+Q&\-398.97575&\-398.90283&72.92\\
DMRG-ENPT2&\-398.97694&\-398.90262&74.32\\
CASPT2&\-398.97755&\-398.90522&72.33\\
\br
\end{tabular}
\end{table}

One may notice that the performance of DMRG-ENPT2 in localized orbitals is not satisfactory. The reason is that SCI and ENPT2 are not orbital invariant \cite{partridge1993ciOrbitalInvariant, murphy1992enpt, malrieu1979ENPTsizeconsistent}. Therefore, the DMRG-ENPT2 based on SCI is not orbital invariant as well.

In \Tref{table2: n2}, we show the energy results by various MC/MR methods based on HF canonical orbitals, localized orbitals (by Pipek-Mezey method) and CASSCF orbitals for both the (nearly) equilibrium and stretched structures of N$_2$. We use the cc-pVQZ basis set and an active space (6e, 6o), and the inter-atomic bond length is $1.1{\rm \AA}$ and $2.0{\rm \AA}$ respectively for the nearly equilibrium and stretched structures. The DMRG-FCI is performed with $M = 1000$.
It is clear that the energies of CASCI, uc-MRCISD, ic-MRCISD and CASPT2 are (nearly) invariant under the rotation from canonical molecular orbitals to localized molecular orbitals, while those by ec-MRCISD+Q and ENPT2 are obviously not, as expected. 
In \Tref{table1: h2s}, it is also reasonably found that the energies of DMRG-ec-MRCISD+Q and DMRG-ENPT2 based on selected CI are not orbital invariant.

\begin{figure}
\centering
\includegraphics[scale=0.5]{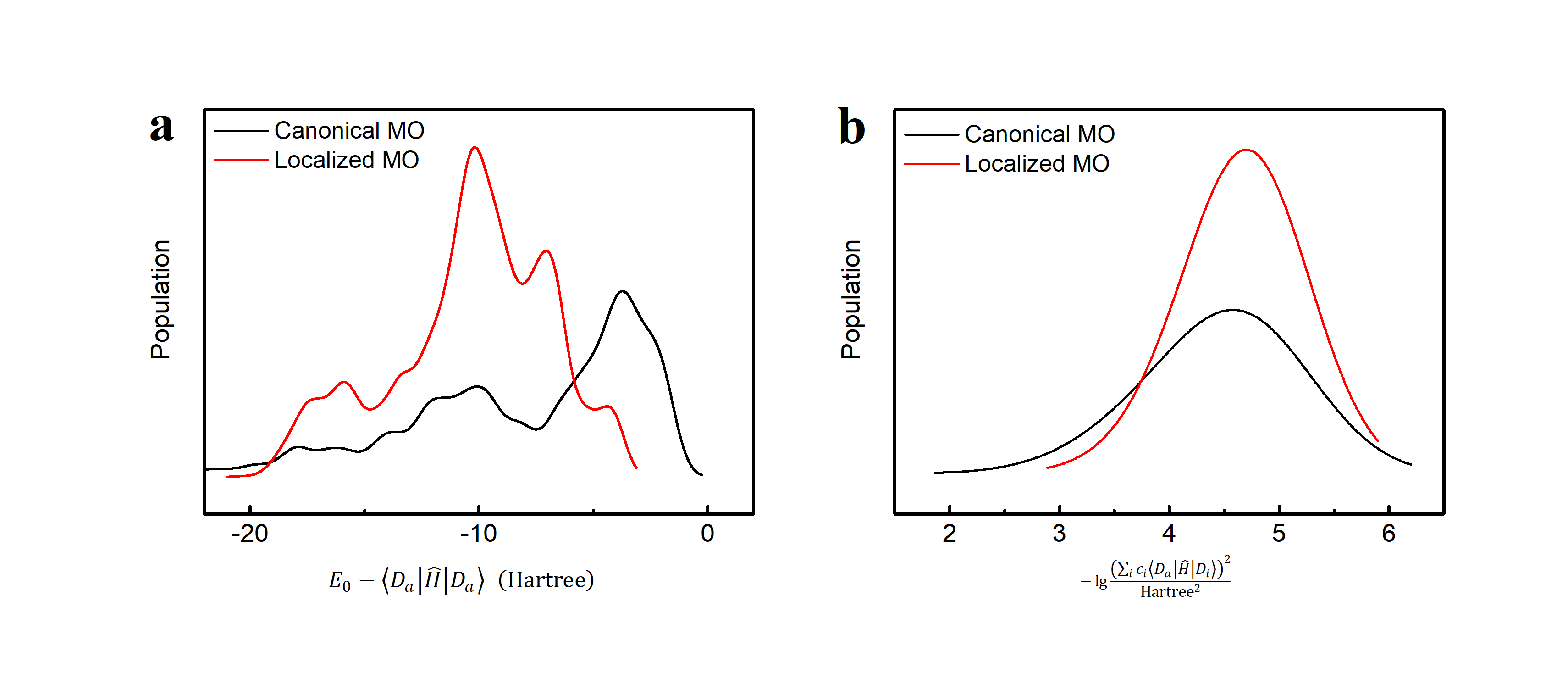}
\caption{Analysis of ENPT2 energy correction for (a) denominators ($E_0 - \langle D_a|\hat{H}|D_a\rangle$) and (b) the negative 10 base logarithm of numerators ($-\lg{\frac{{(\sum_{i\in\Pi}\langle D_a|\hat{H}|D_i\rangle c_i)}^2}{\rm{Hartree}^2}}$).}\label{figure2}
\end{figure}

In fact, the orbital variance of ENPT2 can be easily understood by a simple numerical statistics of the numerator and denominator terms of the ENPT2 energy correction. Here we take N2 with bond length of $1.1{\rm \AA}$ as an example. In \Fref{figure2}, we show the population of denominators ($E_0 - \langle D_a|\hat{H}|D_a\rangle$) and the negative 10 base logarithm of numerators ($-\lg{\frac{{(\sum_{i\in\Pi}\langle D_a|\hat{H}|D_i\rangle c_i)}^2}{\rm{Hartree}^2}}$) by different calculations using canonical and localized molecular orbitals. We can see that the most denominators and numerators of canonical orbitals are bigger than those of localized ones, both of which contribute to the decrease of energy corrections for canonical orbitals. And this leads to the energy decrease in canonical orbitals. The reason is that the delocalization in canonical orbitals lowers the energy of excited configurations and strengthens the couplings between the reference configurations and the excited configurations.

In general, DMRG-ENPT2 performs best in CASSCF orbitals among these three types of orbitals. However, the calculation time of CASSCF is often unbearable. In this case, canonical orbitals is recommended.

\begin{table}
\centering
\caption{Ground state energies of N$_2$ using different MC/MR methods}\label{table2: n2}
\lineup
\begin{tabular}{@{}c|ccc}
\br
N$_2$ (6e, 6o)&$E_{equilibrium}$/Hartree&$E_{stretched}$/Hartree&$\Delta$/mHartree\\
\mr
DMRG-FCI(6e, 106o)&\-109.20028&\-108.87612&324.17\\
\mr
&\multicolumn{3}{c}{Canonical orbitals}\\
\mr
CASCI&\-109.04251&\-108.74500&297.55\\
uc-MRCISD&\-109.19365&\-108.87100&322.65\\
ic-MRCISD&\-109.19313&\-108.87029&322.84\\
ec-MRCISD+Q&\-109.19601&\-108.87313&322.87\\
ENPT2&\-109.22500&\-108.86860&356.40\\
CASPT2&\-109.19203&\-108.88039&311.63\\
\mr
&\multicolumn{3}{c}{Localized orbitals}\\
\mr
CASCI&\-109.04251&\-108.74496&297.55\\
uc-MRCISD&\-109.19365&\-108.87100&322.65\\
ic-MRCISD&\-109.19313&\-108.87029&322.84\\
ec-MRCISD+Q&\-109.15976&\-108.85333&306.43\\
ENPT2&\-109.11431&\-108.81174&302.57\\
CASPT2&\-109.19191&\-108.88039&311.52\\
\mr
&\multicolumn{3}{c}{CASSCF orbitals}\\
\mr
CASCI&\-109.10828&\-108.79861&309.67\\
uc-MRCISD&\-109.19824&\-108.87432&323.93\\
ic-MRCISD&\-109.19809&\-108.87394&324.15\\
ec-MRCISD+Q&\-109.19915&\-108.87473&324.43\\
ENPT2&\-109.20707&\-108.87118&335.90\\
CASPT2&\-109.19506&\-108.87213&322.93\\
\br
\end{tabular}
\end{table}

Besides orbital invariance, there is another important issue for MR methods, size consistency. The SCI based DMRG-ENPT2 is not size-consistent. First, ENPT2 is not size-consistent in delocalized orbitals as previously pointed out by J. P. Malrieu and F. Spiegelmann \cite{malrieu1979ENPTsizeconsistent}. Second, SCI is also well-known to be not size-consistent.

We take a benzene dimer (with a intermolecular distance of $20{\rm \AA}$) as a numerical example to verify. We use the cc-pVDZ basis set and include all valence $\pi$ orbitals in the active space. The SCI, ec-MRCISD+Q and ENPT2 are performed using truncated reference wavefunction constructed via EDGA framework with CI completeness of 0.99999. The results listed in \Tref{table3: benzene} verifies that the ENPT2 method is not size-consistent while CASPT2, SCI and ec-MRCISD+Q are nearly size-consistent when the completeness is nearly equal to 1.

\begin{table}
\centering
\caption{Ground state energies of benzene using different MC/MR methods}\label{table3: benzene}
\lineup
\begin{tabular}{@{}c|ccc}
\br
&$E_{monomer}$/Hartree&$E_{dimer}$/Hartree&$E_{dimer} - 2\times E_{monomer}$/mHartree\\
\mr
CASCI&\-230.77624&\-461.55249&0.00\\
CASPT2&\-230.82016&\-461.64079&\-0.47\\
SCI&\-230.77623&\-461.55247&\-0.02\\
ec-MRCISD+Q&\-230.82713&\-461.65377&0.50\\
ENPT2&\-230.82558&\-461.64630&4.87\\
\br
\end{tabular}
\end{table}

\subsection{Hexacene}

The nature of the ground state of higher acenes is still controversial, so that the S$_0$-T$_1$ energy gap is an important parameter in higher acenes \cite{hajgato2011focal, yang2016nature}. In this work, we take hexacene as an example to calculate its S$_0$-T$_1$ gap. We firstly optimize the S$_0$ and T$_1$ structures of hexacene in $D_{2h}$ symmetry at the B3LYP/6-31G(d) level respectively by using the {\sc GAUSSIAN09} \cite{g09} package. A (26e, 26o) active space, which contains all the valence $\pi$ orbitals and electrons, is then used in our MC/MR calculations with the ANO-L-VTZP and the minimum small atomic natural orbitals basis sets (ANO-S-MB) for C and H respectively. The DMRG-FCI (26e, 126o) calculations with $M=1000$ are performed using HF canonical orbitals. Here, the 126 orbitals include the original 26 active orbitals and the energetically lowest 100 virtual orbitals, corresponding to the same space size of our static and dynamic correlation calculations in DMRG-ec-MRCISD+Q and DMRG-ENPT2. The DMRG-CASCI calculations with $M=1000$ are performed based on the DMRG-CASCI natural orbitals (with $M=500$) generated from the HF canonical orbitals. The SCI, ec-MRCISD+Q and ENPT2 calculations are performed using truncated reference wavefunctions constructed via EDGA framework with CI completeness of 0.99.

The calculated results are listed in \Tref{table4: hexacene}, and the tendency are similar to that of H$_2$S. The DMRG-SCI energy gap differs from DMRG-CASCI by less than 1 mHartree, while the number of reference states for SCI is much smaller than CASCI, no more than 10000. It can be found that the energy gaps of DMRG-ec-MRCISD+Q and DMRG-ENPT2 are in good agreement with each other, and also close to the benchmark DMRG-FCI reference. This verifies the accuracy of DMRG-ENPT2 for general strongly correlated systems (not for very large and extremely correlated ones). Furthermore, the difference of absolute energy values and energy gap between ENPT2 and ec-MRCISD+Q are around 10 and 2 mHartree, respectively. However, we have to note that ENPT2 only uses about 10\% computational time of ec-MRCISD+Q.

So that one can expect that the EDGA-baed DMRG-ENPT2 can be used as an cheap alternative for describing both static and dynamic correlations in large strongly correlated systems.

\begin{table}
\centering
\caption{Ground state energies of and hexacene using different MC/MR methods}\label{table4: hexacene}
\lineup
\begin{tabular}{@{}c|ccc}
\br
hexacene (26e, 26o)&$E_{S_0}$/Hartree&$E_{T_1}$/Hartree&$\Delta$/mHartree\\
\mr
DMRG-FCI (26e,126o)&\-994.44054&\-994.41105&29.49\\
DMRG-CASCI&\-994.34497&\-994.30530&39.66\\
DMRG-SCI&\-994.33237&\-994.29355&38.82\\
DMRG-ec-MRCISD+Q&\-994.42793&\-994.39750&30.44\\
DMRG-ENPT2&\-994.41451&\-994.38323&31.27\\
\br
\end{tabular}
\end{table}

\subsection{H$_{64}$}

\begin{figure}
\centering
\includegraphics[scale=0.2]{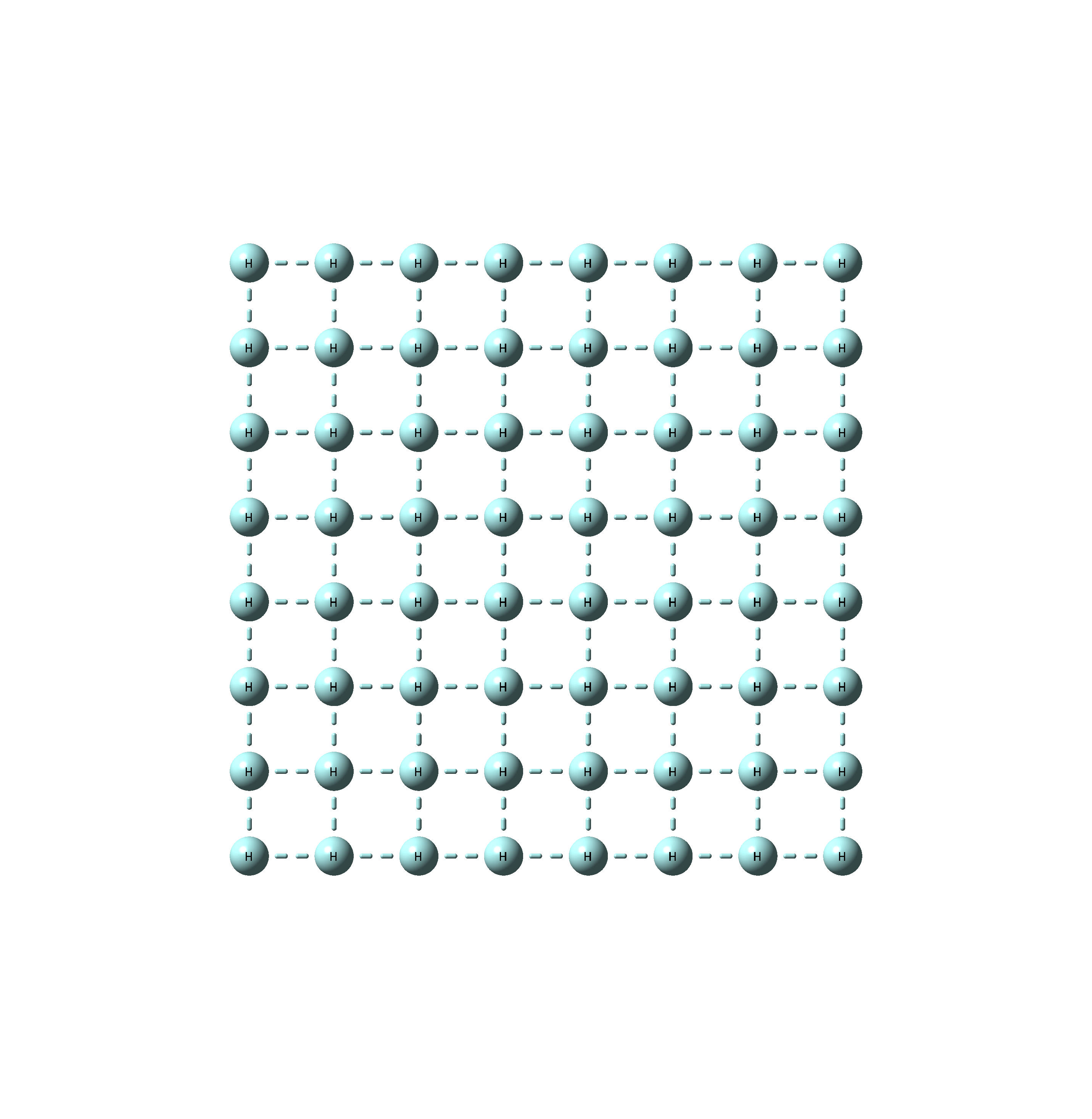}
\caption{Structure of 2D $8 \times 8$ Hydrogen atom square lattice.}\label{figure3.h64}
\end{figure}

\begin{figure}
\centering
\includegraphics[scale=0.5]{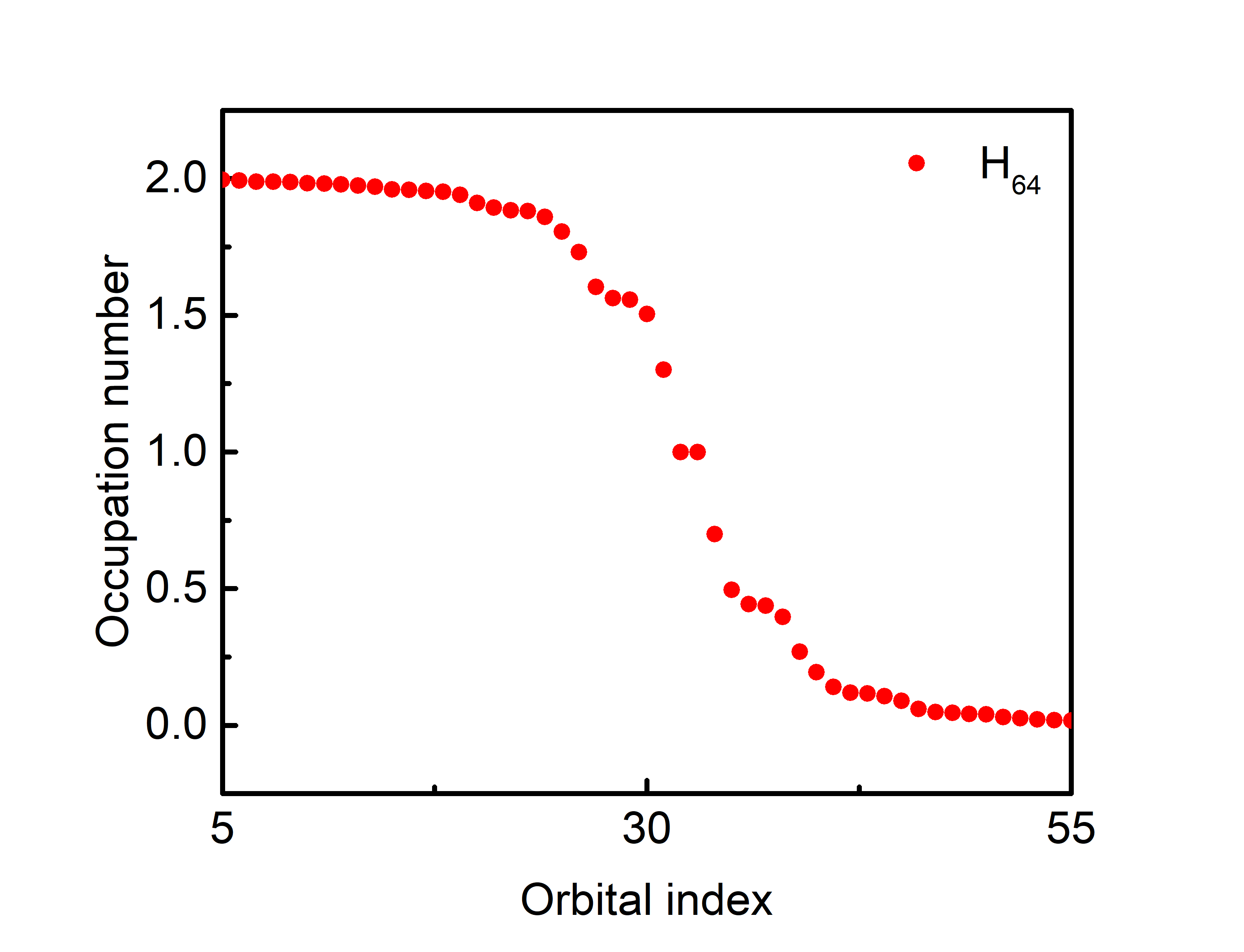}
\caption{Occupation numbers of UHF natural orbitals for 2D $8 \times 8$ Hydrogen atom square lattice.}\label{figure4.on-h64}
\end{figure}

For the well-known strongly correlated system of a 2D $8 \times 8$ Hydrogen atom square lattice (structure in \Fref{figure3.h64}) in which the distance between nearest atoms is 1.3 Å, a preliminary UHF/6-31G calculation is performed firstly. From \Fref{figure4.on-h64} of the occupation numbers of UHF natural orbitals, it is clearly shown that the description of this H64 square lattice requires a large number of active orbitals and there are 42 orbitals with the occupation numbers between 0.02-1.98. We then define a (42e, 42o) active space accordingly and freeze the first 11 orbitals with the occupation numbers higher than 1.98. Next the DMRG-CASCI (42e, 42o) (with $M = 1000$) are performed based on the DMRG-CASCI natural orbitals (with $M = 500$) generated from the UHF natural orbitals. Finally, the SCI and ENPT2 are performed using truncated reference wavefunction constructed via EDGA framework with CI completeness of 0.95. The results are listed in \Tref{table5: h64}. It is shown that DMRG-ENPT can recover the majority of the dynamic correlations outside a large active space by a comparison with the reference DMRG-FCI (42e, 117o) (with $M = 1000$).

\begin{table}
\centering
\caption{Ground state energies of H$_{64}$ using different MC/MR methods}\label{table5: h64}
\lineup
\begin{tabular}{@{}c|c}
\br
H$_{64}$ (42e, 42o)&$E$/Hartree\\
\mr
DMRG-FCI (42e,117o)&\-32.69848\\
DMRG-CASCI&\-32.50450\\
DMRG-SCI&\-32.42406\\
DMRG-ENPT2&\-32.63126\\
\br
\end{tabular}
\end{table}

\section{Conclusion}

In order to account for post-DMRG dynamic electron correlations efficiently, in this work we propose a new method (DMRG-ENPT2) by combining DMRG-SCI with and ENPT2. Test calculations on H$_2$S with an active space of (16e, 15o), hexacene with an active space of (26e, 26o) and 2D H$_{64}$ square lattice with an active space of (42e, 42o) with comparisons to higher level calculations or experimental results show that DMRG-ENPT2 can effectively describe various low-lying spin states in complicated strongly correlated systems. Considering the computational time of DMRG-ENPT2 is only about one tenth of that of DMRG-ec-MRCISD+Q, this provides a cheaper but reliable post-DMRG option for large active spaces beyond 30-40 active orbitals. It should also be noted that DMRG-ENPT2 is not orbital invariant and size-consistent, so DMRG-ENPT2 has to be used carefully.

\section{Conflicts of interest}
There are no conflicts of interest to declare.

\section{Acknowledgments}
The work was supported by the National Key R\&D Program of China (NO.2017YFB0202202), National Natural Science Foundation of China (Grant NOs. 21722302 and 21703260) and the Informationization Program of the Chinese Academy of Science (Grant NO. XXH13506-403).

% ----------> bibliography <----------
\section{Reference}
\bibliographystyle{unsrt}
\bibliography{ENPT.bib}

\clearpage

\end{document}